# Integration of Heterogeneous Modeling Languages via Extensible and Composable Language Components

Arne Haber[1], Markus Look[1], Antonio Navarro Perez[1], Bernhard Rumpe[1], Steven Völkel[2] and Andreas Wortmann[1]

[1]*Software Engineering, RWTH Aachen University, Aachen, Germany*
[2]*Volkswagen Financial Services, Braunschweig, Germany*
`http://www.se-rwth.de`

Abstract: Effective model-driven engineering of complex systems requires to appropriately describe different specific system aspects. To this end, efficient integration of different heterogeneous modeling languages is essential. Modeling language integaration is onerous and requires in-depth conceptual and technical knowledge and effort. Traditional modeling lanugage integration approches require language engineers to compose monolithic language aggregates for a specific task or project. Adapting these aggregates cannot be to different contexts requires vast effort and makes these hardly reusable. This contribution presents a method for the engineering of grammar-based language components that can be independently developed, are syntactically composable, and ultimately reusable. To this end, it introduces the concepts of language aggregation, language embedding, and language inheritance, as well as their realization in the language workbench MontiCore. The result is a generalizable, systematic, and efficient syntax-oriented composition of languages that allows the agile employment of modeling languages efficiently tailored for individual software projects.

## 1 Introduction

Engineering of non-trivial software systems requires reducing the conceptual gap between problem domains and solution domains (France and Rumpe, 2007). Model-driven engineering (MDE) aims at achieving this by raising the level of abstraction from programming of a complete system implementation to abstract modeling of domain and system aspects. In this way, models are raised to be primary development artifacts. Different aspects of complex software systems require different modeling languages to be expressed with. The UML (Object Management Group, 2010), for instance, contains seven structure modeling languages, with class diagrams probably being the most famous, and seven behavior modeling languages as well, e.g., statecharts and activity diagrams. Integration of modeling languages for a software project either requires composing the languages specifically for this project a priori, or designing the independent languages with composition in mind - but without prior assumptions of the actual composition. The former approach yields monolithic language aggregates which are hardly reusable for different projects.

We propose an approach to syntax-oriented black-box integration of grammar-based textual languages developed around the notions of language aggregation, language embedding, and language inheritance (Schindler, 2012; Völkel, 2011). This approach addresses all aspects of syntax-oriented language integration, namely concrete syntax, abstract syntax, symbol tables, and context conditions. Is based on previous work on syntactic modeling language integration (Krahn et al., 2008) and introduces new mechanisms to inter-language model validation. These new mechanisms were briefly introduced at the GEMOC workshop[1] at MODELS 2013 (Look et al., 2013). This contribution explains the concepts and their implementation with the language workbench Monti-Core (Krahn et al., 2010) in detail.

At first, we will motivate the need for language integration in MDE on the example of a cloud-based web system in Sect. 2. Afterwards, Sect. 3 explains the concepts for language integration, and Sect. 4 their support through a language integration framework. Section 5 discusses concepts related to our work. Section 6 concludes this contribution with an outlook on future work and a summary.

---

[1]GEMOC 2013: `http://gemoc.org/gemoc2013/`.

## 2 Motivation

To illustrate our approach we will first motivate the concepts of language aggregation, language embedding, and language inheritance by the example of a cloud-based web system that is described by various, heterogeneous models. The techniques employed in this example will be described in detail in the following sections. Throughout the example, different needs for language integration will arise which we categorize as follows:

- Language aggregation integrates different modeling languages by mutually relating their concepts such that their models can be interpreted together, yet remain independent.

- Language embedding denotes the composition of different modeling languages by embedding concepts of one language into declared extension points of another. Models of the new language thereby contain concepts of both languages.

- Language inheritance is the definition of new languages on the basis of existing languages through reuse and modification of existing language concepts.

Consider a system that receives streams of sensor data from a multitude of internet-connected sensor hardware, e.g., temperature and wattage sensors in buildings, analyzes these streams for patterns, and persists the data into a database. Our aim is to specify this system by using models in a way detailed enough to generate a significant amount of its implementation automatically.

To this end, various system aspects, such as its overall architecture, the data it operates on, and its deployment onto a runtime infrastructure, need to be addressed individually by appropriate modeling languages, including but not limited to architecture models, data models, and deployment models. Yet, many of those aspects are not independent, but mutually related. For instance, for each specified type of data, there are processing logic and database structures specific to it. Consequently, the languages need to be integrated in such a way that their models can reference each other and be interpreted together.

In addition, some aspects of the system are of a more general nature and apply to a wide range of system kinds. Conversely, other aspects are specific to the application domain at hand. For instance, architectural distribution is not only relevant in web systems, but also in many embedded systems, whereas aspects like session management are more specific to web systems. It is desirable to separate the language concepts for general aspects from those for specific aspects in order to facilitate the modularization of languages into reusable language components. Besides, such reuse of existing general languages reduces the need for developers to learn new notations and concepts.

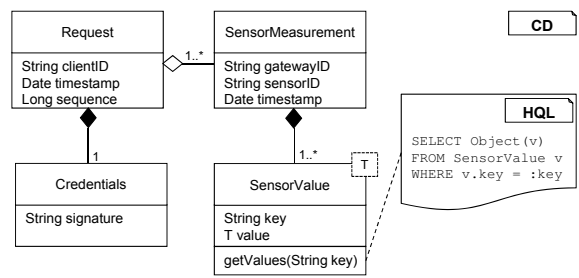

Figure 1: Domain model `SensorData` for sensor measurements.

In the following, we model the domain model and the software architecture of our system using several modeling languages. In doing so, we stress the different needs of language integration that arise from our scenario.

### 2.1 Domain and Data Modeling

Nearly all object-oriented systems operate in the context of a domain model. Such models describe the application's real world context in terms of classes and associations between classes. Their most prominent role is to serve as the basis for the application's fundamental data structures that are used for computation, communication, and state persistence. The latter is typically realized by means of a database that operates according to a database paradigm, such as relational databases or one of various NoSQL flavors. In our example we focus on a relational database which allows for complex queries on data. Such queries are typically expressed in SQL or one of its derivatives, such as Hibernate's HQL (Hibernate website http://hibernate.org/, ), depending on the technology underlying the application's persistence.

Class diagrams (CD) are the foundational modeling language in which domain models are described. We formulate them by means of a textual syntax defined by the UML/P (Rumpe, 2011; Rumpe, 2012; Schindler, 2012), a variant of the UML focused on precise semantics and applicability to generative software engineering. In any case, CDs mainly consist of classes, interfaces, and associations.

Figure 1 shows a graphical representation of the domain model for our example. It consists of classes representing the messages received by the system and the values they convey. Class `SensorValue` is parameterized with a generic parameter T which might be bound with a type defined by another class or by *any type defined by an external language*, e.g., Java. Furthermore, its method `getValues(key)` contains an *embedded HQL expression* that specifies its implementation.

## 2.2 Software Architecture

Cloud-based web systems typically consist of many different components that are distributed over different hardware nodes in a network. In addition, software components and hardware nodes may replicate dynamically at runtime to meet changing system load levels. We model the overall software architecture with a component and connector architecture description language (ADL) (Medvidovic and Taylor, 2000) clArc that is derived from the ADL MontiArc (Haber et al., 2012) and includes cloud-specific language *extensions*. Figure 2 shows the graphical representation of such a software architecture model. It is formulated in terms of hierarchically decomposed components. Components realize the system's functionality by interacting through directed communication channels over which they exchange messages asynchronously. Channels are at each end connected to typed ports which in sum represent a component's interface. These elements are defined by MontiArc.

The example shows two of the extensions clArc introduces. Firstly, components may be marked as replicating, indicating that they depict multiple runtime instances in the actual system. Secondly, components may have service ports in addition to regular ports. These ports represent "vertical" operational interfaces that integrate the component with its runtime environment. In our example, the software architecture model describes a service implemented by the `SensorDataSubmissionHandler` component. This component is decomposed into four interconnected components. The `DataStore` and `EventBroadcaster` component interact with their runtime environment by *referencing* operational interfaces that they request and provide. These interfaces are specified by an external language, e.g., Java. The `PatternMatcher` component uses the language extension of replication to indicate that it may increase or decrease its quantity dynamically at runtime according to system load. Port declarations reference type definitions given by the CD in Fig. 1, thereby denoting that messages exchanged via that port have to correspond to the port's referenced type. Moreover, service ports reference the CD as a whole, indicating that the operational interface represented by that port is made such that it corresponds to basic database operations inferred from the data model interpretation of that CD. It is evident that integration concepts between these different and heterogeneous models are required. In the following section we define such concepts on the language level.

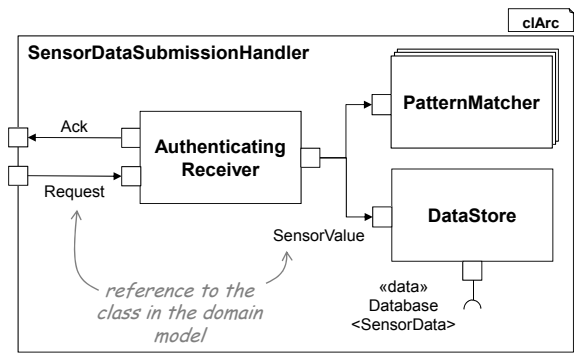

Figure 2: The software architecture model depicting the service `SensorDataSubmissionHandler` which receives and processes streams of sensor data.

## 3 Language Integration Concepts

In (Look et al., 2013) we already gave a first realization of the language integration concepts defined above and outlined their implementation for grammar-based languages in the language workbench MontiCore (Krahn et al., 2010). Here, we describe these concepts as well as their application in detail. In particular, we give a detailed description of the integration concepts, how they manifest in the integrated parsers, the integrated abstract syntax trees (AST), and of references between AST nodes.

The following descriptions make use of the extended grammar format defined by MontiCore. Such grammars serve to systematically derive both the concrete syntax and the abstract syntax of a language, as well as language processing infrastructure such as parsers and pretty-printers. Described briefly, every production of a grammar implies the existence of an AST node class of the same name. The non-terminals of that production form the set of attributes of the AST node class. Their entirety forms the signature of that class. In addition, grammars can define abstract productions and interface productions. Other productions can extend abstract productions and implement interface productions, indicating that their non-terminals can be used anywhere the extended/implemented production's non-terminals are used. Consequently, the resulting AST node classes incorporate the respective extended/implemented signatures into their own. Abstract productions differ from normal productions in that they need to have at least one nor-mal production that extends them. Interface productions work the same way but do not define concrete syntax and hence only consist of non-terminals. An in-depth description of the MontiCore grammar format is given in (Krahn et al., 2010).

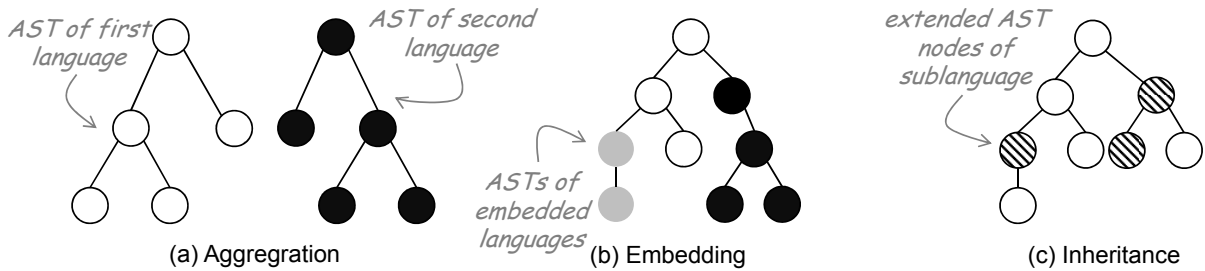

Figure 3: The resulting ASTs for aggregation, embedding, and inheritance. Aggregation results in separate ASTs for each model. Embedding results in a single AST with subtrees embedded at the leaves of the host language. Inheritance results also in a single AST containing extended nodes of the sublanguage.

## 3.1 Language Aggregation

Language aggregation combines multiple languages into a collection of languages (called *language family*), such that models of these languages can be interpreted together but remain formulated in separate artifacts, as shown in Sect. 2.2. There, port declarations reference type definitions given by the class diagram. The individual languages in a language family are loosely coupled and able to mutually reference each other's elements. For instance, a declaration in a model of one language may reference a type declared by a model of another language.

Figure 4 shows how aggregation works on a conceptual level and how concrete aggregations can be defined within the MontiCore framework. The left half shows two grammars (MCG) while the right half shows a class diagram and a clArc model that correspond to their respective grammar on the left. The upper left part shows an excerpt of the grammar for UML/P CDs. In particular, it shows the production of the `CDClass` nonterminal which defines a class as consisting of a modifier, i.e., `private`, `protected`, and `public`, followed by the keyword `class` and the name of the class. An opening curly bracket follows, enclosing arbitrary many attributes, methods or constructors in an arbitrary order, and closing with a curly bracket. The upper right part contains an instance of the production in concrete syntax in which a `SensorValue` class is defined. The lower left part shows an excerpt of the clArc grammar in which the production for a component port declaration is defined. As explained earlier, the `Type` of a port is a name interpreted as a reference to a data type. The lower right part shows an instance of the production referencing a `SensorValue` type.

Via language aggregation CD and ADL models can be combined such that the type references in the ADL model are interpreted as references to class definitions in the CD. The technical realization of language aggregation works in two steps. Firstly, every model of every language is parsed individually, resulting in an AST for each model as shown in Fig. 3. In our example, the type reference in the AST of the architecture model is represented by an AST node of type `Name` containing the name of the reference, whereas the class definition in the AST of the CD is represented by an AST node of type `CDClass`. Secondly, the references are related to each other by a symbol table. Conceptually, the symbol table manages a kind of link between AST nodes. Technically, the links are implemented by adapter classes to allow for flexible linking to AST nodes from other languages. The details of this are described in Sect. 4.

Language aggregation is adequate for modeling different aspects of a system, each of which can be understood on their own. Each aspect is then described by individual model documents in specialized modeling languages. Through aggregation, these models are related to each other without infringing a tight coupling between them. Thereby, models can be reused in different combinations and in a modular way. For instance, the same CD can be used to define the same types which can then be referenced by different models and in different development projects.

## 3.2 Language Embedding

Language embedding combines languages such that they can be used in a single model. To this end, an embedding language incorporates elements from other languages at distinguished extension points. Even though this gives the impression of tight coupling, the individual languages are still developed independently and integrated in a black-box way. References between elements of embedding and embedded lan-guages work similarly to language aggregation. Sect. 2.1 shows an example of CDs with embedded HQL.

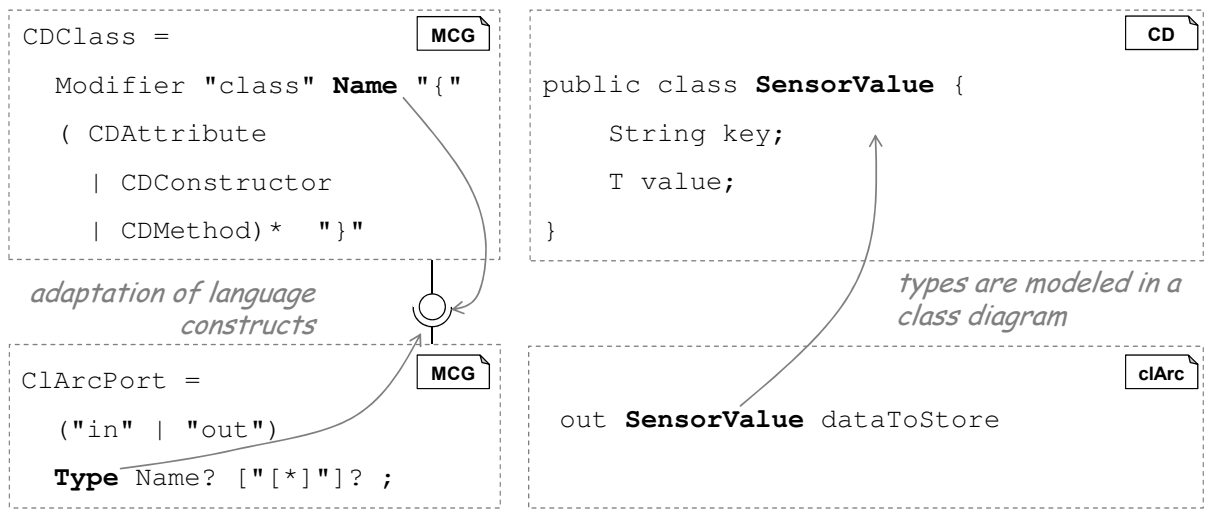

Figure 4: The mechanism for aggregating languages. By adaptation between elements of two independent languages referencing between them is achieved. The right half of the figure shows concrete models of aggregated languages referencing each other.

Figure 5 shows how the embedding is accomplished within the MontiCore framework. The upper left part again shows an excerpt of the CD grammar with the production of the nonterminal `CDMethod`. The production describes a method with a modifier, a return type, a name followed by parameters enclosed in round brackets, and a `Body`. The body is defined as an external nonterminal. Such external productions act as the extensions points into which elements from other languages can be embedded. In fact, the language is not complete as long as its external productions have not been bound to external language elements. Note that neither the grammar nor the external production contain any information about filling the external nonterminal, leaving the binding to a later stage. The lower left part shows an excerpt of a HQL grammar containing the definition of a HQL block statement that encloses multiple statements with curly brackets. The `HQLStatement` nonterminal is production comprises further nonterminals, such as `SELECT` or `INSERT` statements. Again, the HQL grammar does not contain any explicit reference to language embedding. The actual embedding is specified in the language configuration model, shown in the middle part of Fig. 5. The model maps the nonterminal `HQLBlock` to the external nonterminal `Body`. It is also possible to embed several languages into a single external production by mapping several external non-terminals to it, for instance, vanilla SQL. After parsing, the resulting AST consists of different nodes of the different languages, as shown in Fig. 3. Nodes from embedded languages manifest as subtrees attached in place of the node representing the external production.

Language embedding is especially useful when the language developer does not want to force the use of a specific language but allows to choose the sublanguage later. For instance, it can be used to embed different action languages within a structural language to specify behavior.

### 3.3 Language Inheritance

Language inheritance can be used to extend or refine an existing language. For this purpose, MontiCore allows to define new languages on the basis of existing languages by reusing, modifying and overriding their productions. The example in Sect. 2.2 illustrates how the clArc language extends the MontiArc language and adds cloud-specific extensions.

Figure 6 illustrates how this extension is defined within MontiCore's grammar format. The upper left part shows a production (with details omitted) from the MontiArc grammar which specifies the nonterminal `ArcPort`. Each port starts with either the keyword `in` or the keyword `out` followed by a `Type` and a `Name`. Both `Type` and `Name` define possible identifiers for types and instances names similar to the naming scheme used in Java. The upper right part shows a model element that conforms to the production shown in the upper left part. The lower left part shows an excerpt of the textual clArc grammar extending the MontiArc grammar. The name of the grammar is followed by the keyword `extends` and a reference to the extended grammar. The production of the nontermi-nal `ClArcPort` also contains the keyword `extends` and a reference to the name of the extended production inherited from the MontiArc grammar. The left-hand side of the production contains all elements present in the parent production and adds the possibility of specifying an additional terminal `[*]` which denotes replicating ports. It is not obligatory to keep all elements of a production that have been present in the parent production. Instead, it is also possible to leave some out, reorder them, add new elements in between, or even remove all elements. The lower right part shows a model element that corresponds to the production on the lower left.

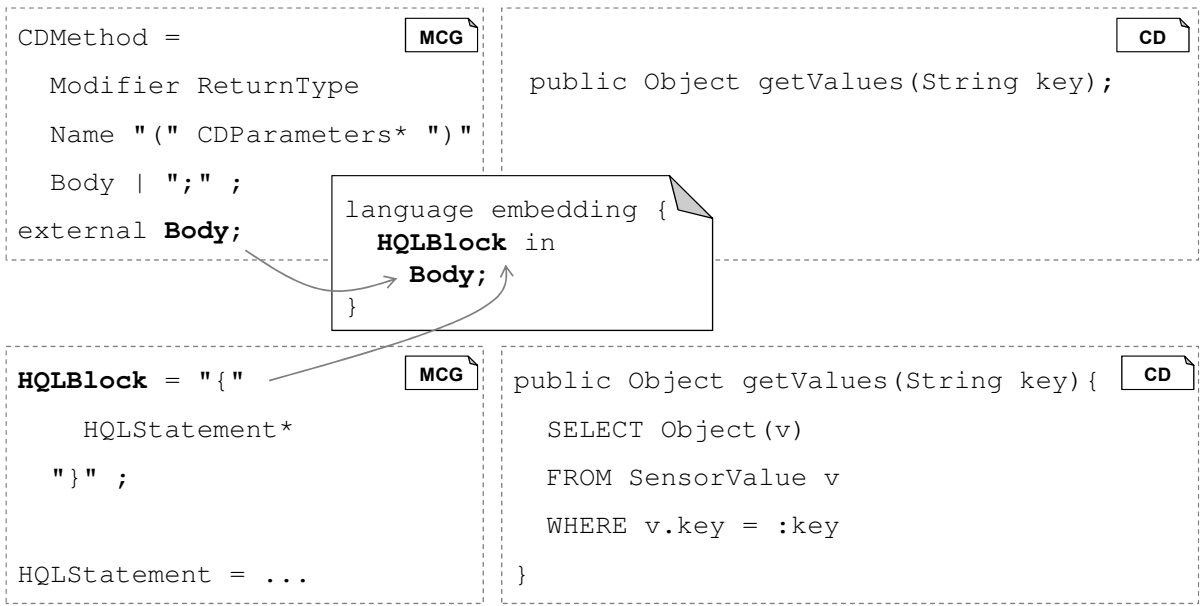

Figure 5: The mechanism for embedding languages. By declaring an external nonterminal `Body` and a separate mapping artifact, nonterminals of arbitrary languages, such as `BlockStatement` can be embedded. The right half of the figure shows a concrete model with the embedded element.

Productions of extended languages (or "parent" language) are "virtually'" copied into the extending new language where they can be referenced from new productions. In addition, new productions can individually extend productions from the parent language and thereby inherit that production's interface. This means that the extending production can be used anywhere the non-terminal from the extended production is used. The resulting new AST nodes consequently implement the signatures of their parent counterparts.

The right part in Fig. 3 illustrates the structure of ASTs from inheriting languages. The generated parser for the sublanguage is able to parse text corresponding to the parent language as well as text corresponding to the sublanguage, and consequently creates an AST containing node types from both languages. Since parent grammars and nodes are referenced by names, name collisions can occur. To prevent this, the language designer may use full qualified names formed by the respective grammar's package, the grammar's name and the name of the production.

Language inheritance is particularly useful for reusing existing concepts of languages while extending them with new concepts. It is applicable when the inheriting language is conceptually similar to the parent language.

## 4 Language Integration Framework

In this section, we describe a framework to (1) implement *symbol table infrastructures* that relate model elements in composed languages in a black-box way and to (2) configure them for concrete language compositions based on the concepts introduced above.

Section 3 described how language embedding and language inheritance manifest in the ASTs of processed models that are instances of combined languages (cf. Fig. 3). Both mechanisms make information about embedded or inherited model elements directly available in the AST for further model processing, such as code generation. However, this is not the case for models of aggregated languages in which elements of one language reference elements of another language by name. Here, the AST nodes only contain the raw name of the referenced model element. The same holds for embedding and embed-ded languages which refer to each other through raw names as well. Consequently, additional infrastructure is necessary to translate raw name references into information about referenced model elements.

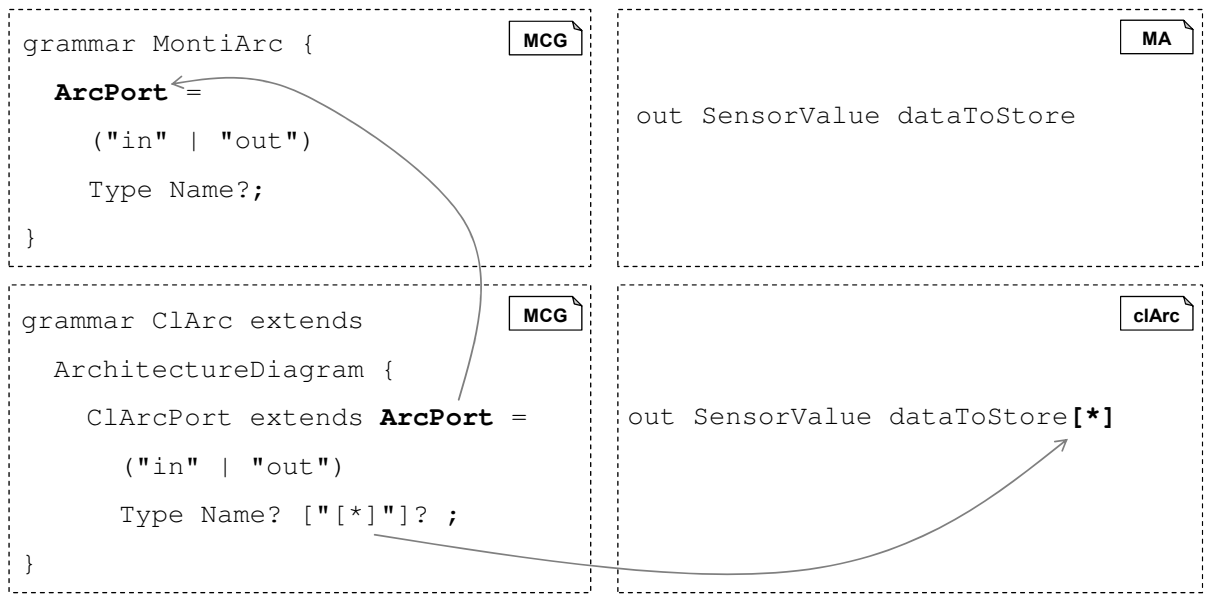

Figure 6: The mechanism for inheritance between languages. By declaring an extension, languages can inherit from each other and are able to override productions. The right half of the figure shows a concrete model using the inheritance.

In the following, we describe an infrastructure named symbol table that a) allows to acquire information from referenced models as well as b) to transparently interpret elements of one language as elements of another. Compared to traditional symbol table techniques, our realization must be able to translate these different kinds of concepts between languages. For example, automata know about states and input signals, whereas Java knows nothing about these. To integrate these languages nonetheless, the concept of a state must be translated into Java in a meaningful way. For instance, states could be mapped to an enumeration or also to subclasses (as, e.g., in the state pattern (Gamma et al., 1995)). To keep both languages, independent, we cannot define this translation in either language. Instead, we need to define it as a separate artifact during language integration. The ability to do this is one key feature of our symbol table framework.

## 4.1 Symbol Table Concepts

A *symbol table* is a data structure that is used to store and to resolve identifiers within a language. An identifier, such as a name, is associated with further information from the corresponding language element. This way detailed information may be gathered from the symbol table by resolving an element using its name. In (Völkel, 2011) the most important parts of a symbol table are defined. These are:

**Definition 1** (Entry and Kind). *A symbol is an entry in a symbol table that represents a named element of a model. It has a well defined signature determined by its kind. This structure allows to store and read kind specific information together with the entry.*

An entry may be in distinct states that represent the completeness of its name and the associated information. For an *unqualified* entry only its unqualified name, e.g., SensorValue is known, if the full qualified name, e.g., de.se.SensorValue is known the entry is in the *qualified* state. If an entry is in the *full* state, further information respectively entries, e.g., the methods of type de.se.SensorValue, are associated to the entry.

**Definition 2** (Scope). *A scope of an entry is that part of a model in which the element that is represented by the entry may be referenced by its name. Thus, the entry is visible and not hidden within its scope.*

**Definition 3** (Namespace). *A namespace is a section of a model in which names and corresponding entries are managed together in symbol tables. Usually, namespaces are attached to nonterminals that open namespaces and are thus organized hierarchically. A namespace may import names and corresponding entries from other namespaces, e.g., from a parent namespace. Entries with the same name stored in different namespaces may shadow each other, if the namespaces are related.*

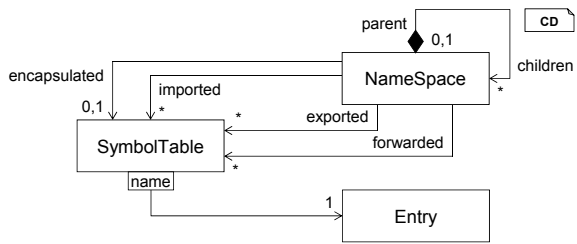

Figure 7: Hierarchical namespaces, their symbol tables and related entries (cf. (Völkel, 2011)).

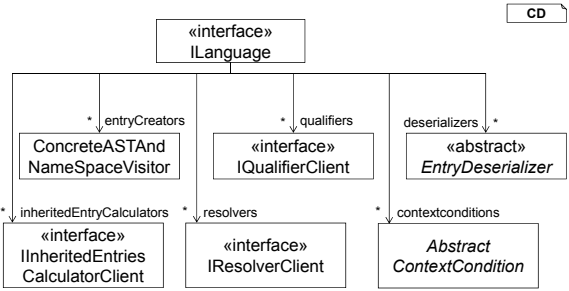

Figure 8: Technical components of a symbol table (cf. (Völkel, 2011)).

The structural relation of these elements is depicted in Fig. 7. A `NameSpace` may have arbitrary many child namespaces and an optional parent namespace to represent a namespace hierarchy. A `SymbolTable` associates entries (`Entry`) with their name. Entries are serialized in a preprocessed, condensed form that allows for fast loading of its contained information.

## 4.2 Symbol Table Components

The previously described symbol table and namespace structure has to be created for each language. The MontiCore framework (Grönniger et al., 2006) provides infrastructure for a uniform development of technical symbol table components for modeling languages. The most important classes and interfaces are depicted in Fig. 8.

Each concrete modular modeling language is presented by the interface `ILanguage`. It offers the technical components needed to create namespaces and symbol tables for an instance of that language. Its entry creators (subclasses of `ConcreteASTAndNameSpaceVisitor`) are used to set up the namespace hierarchy of a model. Then, entries for model elements are created and organized in the symbol tables of the given namespace hierarchy. The registered `IInheritedEntriesCalculatorClients` are used to compute if entries from imported namespaces are hidden by locally defined entries. An

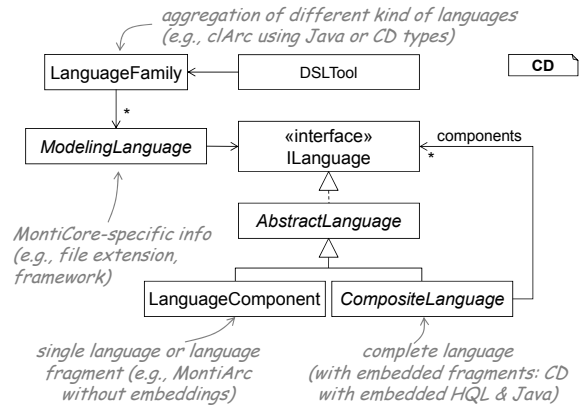

Figure 9: Technical realization of MontiCore's language composition mechanisms (cf. (Schindler, 2012; Völkel, 2011)).

`IQualifierClient` has to be provided for each element kind of a language which instances may be referenced within the current or another model, such as a referenced type of a field. The concrete qualifier client is used to transfer entries with the corresponding kind from the unqualified to the qualified state. Resolver clients (`IResolverClient`) have to be provided for each entry kind that may be referenced within the current namespace hierarchy. The registered deserializers (`EntryDeserializer`) load serialized entries from externally referenced models. They are used to transition the entries that represent a referenced model element from the qualified to the full entry state. Associated context conditions that extend the abstract class `AbstractContextCondition` are used to check if processed models are well formed.

This way a concrete `ILanguage` module offers all means to process models or model parts of a certain language and produce a corresponding namespace and symbol table hierarchy. The provided infrastructure additionally alleviates inter-model relations that allow to resolve external information defined in related models. How to combine these components in several ways to realize the language integration concepts presented in Sect. 3 is described in the following subsection.

## 4.3 Configuration of Language Compositions

Language integration requires different effort depending on the type of integration. The composition takes place hierarchically to enable application of mechanisms in the best possible order. Fig. 9 shows the different language concepts required to achieve this compositionality.

Language aggregation of two or more existing modeling languages is implemented in `LanguageFamily` instances. These gather the different independent `ModelingLanguage`s together with the inter-language infrastructure, such as resolvers, qualifiers and adapters for symbol table integration, as well as factories and inter-language context conditions. Language families are used by MontiCore's `DSLTools` to process sets of heterogeneous but related models. The clArc language family, for instance, comprises a modeling language for architecture models and a modeling language for CDs.

A `ModelingLanguage` is a black-box language and contains language-specific information such as the file ending. It may contain either a single language or a composition of embedded languages, such as CDs with embedded HQL. Therefore, modeling languages contain a hierarchy of `ILanguage` interfaces. Based on this, MontiCore creates the infrastructure to parse model instances accordingly. This infrastructure contains the correct combination of parser and lexer for the model at hand.

For single languages, modeling languages contain only a single `LanguageComponent` which contains the symbol table infrastructure and context conditions necessary. A `LanguageComponent` contains the information required to process symbols of the respective single language, i.e., how entries are created, deserialized, qualified, resolved, which context conditions are available, and which entry types are exported (see Sect. 4.2). For embedded languages, modeling languages contain a hierarchy of `CompositeLanguage`s. These are composed of the embedded languages that themselves are represented by an implementation of `ILanguage`. Language components and composite languages are implemented as `AbstractLanguage` which provide common functionality used by language components and composite languages. They implement the interface `ILanguage` to allow utilization in a composite (Gamma et al., 1995). Using the composite pattern for embedding allows to reuse the resulting language combinations easily in different contexts, e.g., to embed Java and HQL into CD. Composite languages and language families can be considered as the glue between languages and their symbol tables as they hold the required adapters, resolvers, qualifiers, and context conditions for their specific composition.

Figure 9 also shows, that the order of language aggregation is arbitrary and depends on the language engineer. Whether a subset of the embedded languages should define a new `ModelingLanguage` solely depends on the desire to reuse this combination. It might, for example, be useful to combine CD and HQL first, and to reuse the resulting combination with different action languages. Please note that language inheritance is not reflected in this structure, as the resulting combined abstract syntax does not necessarily require any interaction on symbol level. If this however is necessary, usually a new 'main' entry has to be created which contains the additional information resulting from the language extension. Thus clArc introduces a new component entry to contain the new language features. To be reusable with existing language integration infrastructure of the inherited language, these need to be adapted to its entries accordingly. Hence, clArc component entries need to be adapted to MontiArc component entries. Similar to the development of the symbol table of a new language, entries, entry creators, qualifiers, and resolvers have to be registered for elements added by the inheriting language. The inheriting language can also reuse context conditions of the inherited language and add new ones.

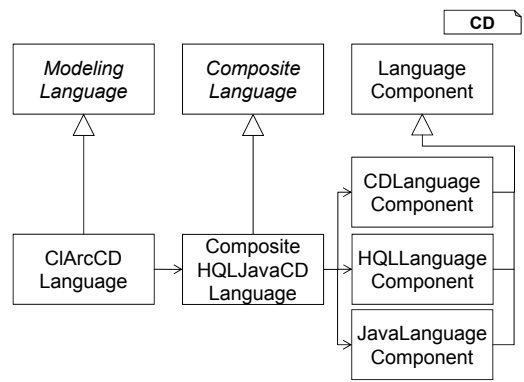

Figure 10: Language composition for the UML/P `ClarcCDLanguage` (cf. (Schindler, 2012)).

Figure 10 illustrates the language composition mechanisms on the clArc CD language `ClArcCDLanguage`. The language allows to model CDs with embedded Java and HQL as illustrated in Sect. 2.1. As this embedding happens on the concrete syntax, there is no need to reference models of other languages by name. Therefore, the language is implemented as a modeling language that contains the composite language `CompositeHQLJavaCDLangauge` which realizes the embedding. `CompositeHQLJavaCDLangauge` is composed of language components for CD, Java, and HQL respectively, and contains adapters between the three languages as well as inter-language context conditions. Language-internal context conditions are defined in the language component. Cross-language inter-model context conditions are defined in language families, whereas cross-language intra-model context conditions are defined in composite languages. Adaptation between entries of two embedded languages, such as the types of embedded HQL and the embedding CD, requires the composite language to provide adapters, qualifiers, and resolvers for type pairs. Adaptation between aggregated languages of a language family requires to configure these elements in the language family instead. Figure 11 shows the elements required to adapt type entries of a HQL language to type entries of a CD language. Integration requires to provide a new adapter factory marked responsible to create entries of a certain type (here HQL type entries for CD types). When a CD type needs to be qualified or resolved for embedded HQL, the factory produces an adapter which behaves like a HQL type entry, but delegates all methods to the adapted CD type entry.

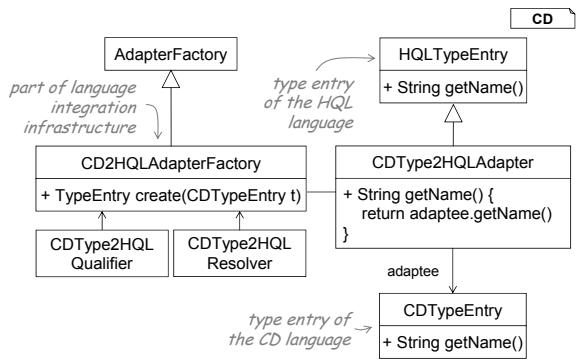

Figure 11: Adaptation between type entries of HQL and those of the CD language.

Using adaptation on the levels of composite languages and language families allows to develop languages without consideration of a posteriori integration. As the languages are free from integration premises, they can be composed arbitrarily.

## 5 Related Work

We have presented three mechanisms for the integration of modeling languages. Integration takes place on the syntactical level and enables language aggregation, language embedding, and language inheritance. Related to our contribution are other studies and approaches on general syntax-oriented language integration. We do neither discuss language integration for specific language families (Barja et al., 1994; Groenewegen and Visser, 2008) as these are usually specifically created to be integrated, nor do we discuss semantic language integration (Grönniger and Rumpe, 2011; Hedin and Magnusson, 2003; Wende et al., 2010; Wyk et al., 2008).

A study on language composition mechanisms distinguishes the mechanisms: "language extension, language restriction, language unification, self-extension, and extension composition" (Erdweg et al., 2012). The authors' notion of language extension also requires that languages can be composed a-posteriori and distinguishes language extension from language integration. Our approach to language extension allows to overwrite nonterminals from the extended language in order to reduce expressiveness. The proposed notion of "language unification" matches our definition of language aggregation, where two independent languages can be used "unchanged by adding glue code only". In their definition of "self-extension", the authors start from a different definition of "language embedding" than we do: there, language embedding is that a "domain-specific language is embedded into a host language by providing a host-language program that encapsulates the domain-specific concepts and functionality" (Hudak, 1998) which defines the use of domain-specific programs and is hardly recognizable as language embedding. Accordingly, the author's definition of "self-extension" requires that "the language can be extended by programs of the language itself while reusing the language's implementation unchanged"–which also allows to "embed" languages as strings into the host language, e.g., SQL queries or regular expressions in Java. MontiCore does not provide an explicit "self-extension" mechanism, but supports it by embedding action languages allowing definition of programs. MontiCore languages also support the language extension composition mechanisms denoted as incremental extension and language unification as defined in (Erdweg et al., 2012).

A recent contribution on language workbenches (Erdweg et al., 2013) provides an overview of existing tools and their features. In particular, Ensō (van der Storm et al., 2014), Más (Más website http://www.mas-wb.com, ), MetaEdit+ (Kelly et al., 1996), MPS (Dmitriev, 2004), Onion (Erdweg et al., 2013), Rascal (Klint et al., 2009), Spoofax (Kats and Visser, 2010), SugarJ (Erdweg et al., 2011), the Whole Platform (Solmi, 2005) and XText (Eysholdt and Behrens, 2010) are reviewed. This review considers four dimensions: syntax, validation, semantics and editor services. According to the overview all of the presented tools are able to achieve syntactical composition via different mechanisms. Nevertheless the composition on the validation depends on the validation features of the respective workbench. Only MPS, SugarJ and XText provide validation for naming and type checking similar to our approach to syntax-oriented language integration, namely concrete syntax, abstract syntax, symbol tables, and context conditions. In (Voelter, 2013) the composition of languages in MPS is shown in more detail. To compose types new type definition rules have to be applied to infer types via unification. Since MontiCore is not projectional and uses independent parsers we define these connections between AST elements via our adaptation mechanisms and not via generic type definition rules. Furthermore (Voelter, 2013) distinguishes between language combination, extension, reuse and embedding. While language combination and reuse are similar to our notion of aggregation, language extension corresponds to our language inheritance, and language embedding is congruent with our concept of embedding.

A recent contribution on model-based language integration (Tomassetti et al., 2013) highlights cross-language context conditions as an important source for errors. The authors propose to develop reusable cross-language context conditions and sketch how these can be implemented with their language workbench. It remains to be discussed how context conditions checking semantic properties specific to a language family can be designed for reuse.

Another approach to deal with the complexity of language integration is to employ domain-specific embedded languages (DSELs) in a host language (e.g., Scala) (Hofer and Ostermann, 2010). Regarding our example, this circumvents the problems arising from using data types between languages and allows to reuse existing development infrastructure. These approaches focus on syntax-oriented integration as well, but language reuse is limited to languages of the same host language and often DSELs lack explicit meta-models usable for integration purposes.

Attribute grammars (Knuth, 1968) allow to enrich grammar symbols with computation rules. Research in attribute grammars led to promising results regarding language integration, like Forwarding (Wyk et al., 2002), and produced capable language workbenches as well (Wyk et al., 2008). Using multiple inheritance of attribute grammars to integrate is another interesting approach (Mernik, 2013) to language integration which suffices to fulfill the language composition mechanisms identified in (Erdweg et al., 2012).

## 6  Future Work and Conclusion

Engineering of complex software systems requires MDE where language integration can help to deal with the heterogeneity of modeling languages involved. We introduced language aggregation, language embedding, and language inheritance by example. These language integration techniques allow integration of languages without stipulating possible integration partners or mechanisms a priori. This enables to compose languages with minimal effort.

We have illustrated how these integration mechanisms are implemented in MontiCore. To achieve cross-language resolution of names, the symbol table, language families, and modeling languages were introduced. Language families and modeling languages contain the glue to enable cross-language model usage. This glue is implemented in the form of adapters between entries of symbol tables.

In the future we will examine whether parts of symbol table infrastructure can be generated from the language and its models directly. We further will investigate whether the different language integration definition mechanisms (e.g., grammars for embedding, symbol table for aggregation) can be unified. Furthermore, it might be possible to generate the cross language infrastructure from enriched models of the respective language as well.